\documentclass[10pt,letterpaper]{article}
\usepackage{opex3}
\usepackage{amsmath}
\usepackage{graphicx}
\usepackage{subfig}

\begin{document}

\title{Manipulating light at distance\\by a metasurface using momentum transformation}

\author{Mohamed~A.~Salem$^{*}$ and Chrsitophe~Caloz}

\address{Department of Electrical Engineering, \'{E}cole Polytechnique de Montr\'{e}al,\\Montr\'{e}al, Qu\'{e}bec H2T 1J3, Canada}

\email{*mohamed.salem@polymtl.ca} 



\begin{abstract}
A momentum conservation approach is introduced to manipulate light at distance using metasurfaces. Given a specified field existing on one side of the metasurface and specified desired field transmitted from the opposite side, a general momentum boundary condition is established, which determines the amplitude, phase and polarization transformation to be induced by the metasurface. This approach, named momentum transformation, enables a systematic way to synthesize metasurfaces with complete control over the reflected and transmitted fields. Several synthesis illustrative examples are provided: a vortex hypergeometric-Gaussian beam and a ``delayed-start'' accelerated beam for Fresnel region manipulation, and a pencil beam radiator and a holographic repeater for Frauenhofer region manipulation.
\end{abstract}


%
%

\section{Introduction}

Manipulating light at a distance away from its source is a key requirement in advancing light wave technologies, such as communication, imaging, and sensing. To enable such advances, various approaches are being devised, with an increasing interest in metamaterial structures (see e.g.,~\cite{Capolino:09} for an historical overview). Most of the work reported on light manipulation using metamaterial structures assumes local manipulation (field is controlled inside the metamaterial or in its direct vicinity) such transformation optics~\cite{Pendry:06,Liu:13} and conformal mapping techniques~\cite{Leonhardt:06} to perform operations that are usually not possible using conventional materials, such as analog computation~\cite{Silva:14}. Recently, two-dimensional metamaterial surfaces, generally named metasurfaces, were introduced to shape wavefronts beyond what was possible using frequency selective surfaces. These metasurfaces found applications in the generalized law of refraction~\cite{Yu:11}, wavefront shaping virtual Huygens sources~\cite{Pfeiffer:13a,Pfeiffer:13b,Selvanayagam:13}, and computational imaging~\cite{Hunt:13}.

In this work, the foundation for a general framework is established to manipulate light waves using metasurfaces. The proposed approach is based on the concept of light momentum conservation, and is therefor named the momentum transformation method. This method, although straightforward, presents a rigorous and systematic approach to manipulate light at distance with unprecedented flexibility and insight and, to the best of the authors' knowledge, is not previously reported in the literature. In addition, the presented method provides the initial guidelines for metasurface synthesis with a physical insight into the light wave behavior.

The paper is organized as follows: Section~\ref{sec:2} derives the momentum transformation method, starting with the scalar case in Sec.~\ref{sec:2.1}, and generalizing to the vectorial case, using momentum space expansion vector basis functions, in~\ref{sec:2.2}. Section~\ref{sec:3} provides some guidelines for the metasurface realization in terms of its scattering elements in the momentum space. Several illustrative examples for near-field and far-field manipulations are presented in Sec.~\ref{sec:4.1} and Sec.~\ref{sec:4.2}, respectively. Finally, Sec.~\ref{sec:5} concludes the paper.

\section{The Momentum Transformation Method}
\label{sec:2}

Since the momentum (or the wavevector) of light waves is conserved in homogenous media, a forced change in any of its components necessarily affects the other components in order to ensure total momentum conservation. Hence, forcing local momentum changes in the $k$-space, should allow one, in principle, to reshape the light wave in a controlled fashion. The interaction between the metasurface and the fields is local in space, i.e., the difference between the field at a certain point one side of the metasurface and the same point on the opposite side is solely determined by the metasurface transfer function at the same point. Formal mathematical derivation is given in the appendix. The local interaction in the direct space translates into a convolution relation in the momentum space. Although $k$-space manipulation is convolutional in nature, it allows for superior light manipulation at distance compared to direct space manipulation, as will be shown. Metasurfaces thus emerge as best suited candidates for the momentum transformation since they can effectively manipulate the transverse components of the wavevector.

The following sub-sections establish two approaches for momentum transformation. The first approach deals with scalar fields, such as acoustic fields and paraxially approximate fields encountered in optics. The second approach deals with vectorial fields and is therefore relevant to electromagnetic waves manipulation.

\subsection{Scalar Transform}
\label{sec:2.1}

\begin{figure}[tbp]
\centering
\includegraphics[width=2.5in]{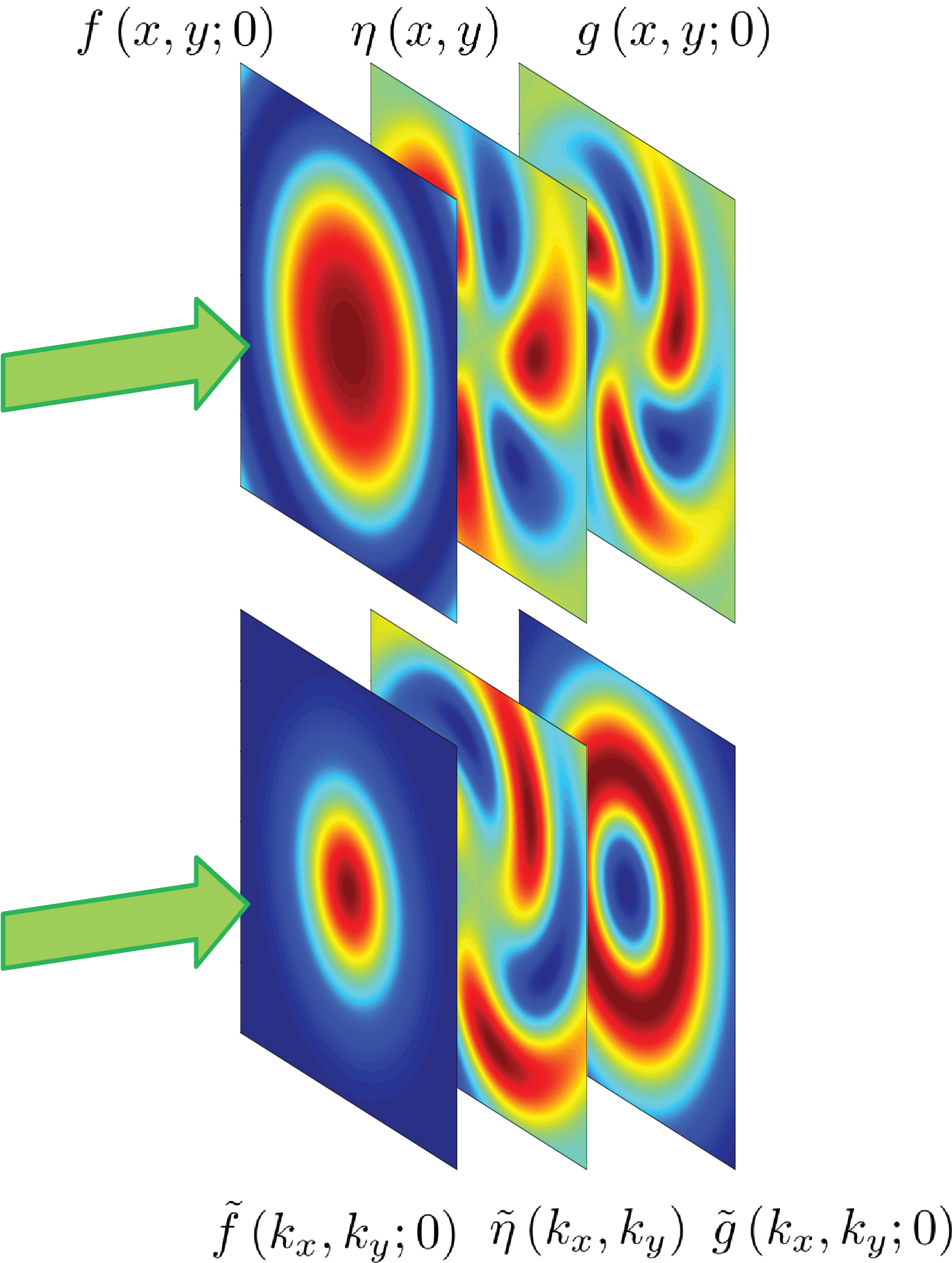}
\caption{Illustration of the interaction between an impinging wave field and the metasurface resulting in a transformed transmitted field. In the upper part, the direct space representation of the impinging field $f(x,y;0)$, the transmitted field $g(x,y;0)$, and the metasurface function $\eta(x,y)$ are plotted. In the lower part, the corresponding momentum space quantities are plotted.}
\label{fig:1}
\end{figure}

In the scalar case for an infinitely thin metasurface, the interaction between the fields and the metasurface is local, i.e., linear-shift variant. Thus, for a metasurface located at $z=z'$ in the Cartesian coordinate system with the fields $f(x,y)$ at $z=z'^{-}$ and $g(x,y)$ at $z=z'^{+}$, the linear-shift variant relation reads%
\begin{align}
g\left(x,y\right) &= \iint_S { f\left(x',y'\right) h\left(x',y'; x,y\right) dx' dy'}, \nonumber \\
&= \iint_S { f\left(x',y'\right) \delta\left(x'-x,y'-y\right) \eta\left(x,y\right) dx' dy'}, \nonumber \\
&= f\left(x,y\right) \eta \left(x,y\right), \label{eq:lsv}
\end{align}%
where $z'$ is suppressed, $S$ denotes the plane of the metasurface and $h(x',y';x,y)$ is the generalized transfer function of the metasurface, which reduces to $\eta(x,y)$ due to local interaction. 

The fundamental transformation between the direct space and the momentum space in the vicinity of the metasurface plane is established through the two-dimensional Fourier transform pair,%
\begin{align}
\tilde{f}\left(k_x, k_y;z'\right) &= \frac{1}{2\pi} \iint_{-\infty}^{\infty}{f\left(x, y;z'\right) e^{-i \left[ k_x x + k_y y \right] }d x d y}, \label{eq:meta} \\
f\left(x, y;z'\right) &= \frac{1}{2\pi} \iint_{-\infty}^{\infty}{\tilde{f}\left(k_x, k_y; z'\right) e^{i \left[ k_x x + k_y y \right]} d k_x d k_y}, \nonumber
\end{align}%
where the tilde represents momentum space functions, and $(k_x, k_y)$ are the transverse components of the wavevector, $\mathbf{k}$. Relation~(\ref{eq:lsv}) becomes a convolutional relation in the $k$-space reading $\tilde{g} = \tilde{f} \ast \tilde{\eta}$. Alternatively, this relation could be written to express the metasurface function explicitly as%
\begin{equation}
\label{eq:scalar}
\tilde{\eta}\left(k_x,k_y\right) = \tilde{g}\left(k_x,k_y\right) \ast \tilde{\zeta}\left(k_x,k_y\right)
\end{equation}%
where $\zeta = 1/f$. This momentum relation is illustrated in Fig.~\ref{fig:1}.

Equation~\eqref{eq:scalar} by itself does not offer much advantage over its direct space counterpart~\eqref{eq:lsv}, which is commonly used in optical applications. However, it could be argued though that the momentum space representation may offer a better insight into the dynamic range of momentum variation (upper and lower bounds) to be induced by the metasurface, and hence on the possibility to synthesize the metasurface with electromagnetic scattering elements. Moreover, the actual advantage of the momentum space representation is evident when the transmitted field is required to be manipulated away from the metasurface. In order to manipulate the transmitted field at a distance $z=d$ away from the metasurface plane, a reverse propagation operator is introduced. The role of the reverse propagator is to propagate the $k$-space components of $\tilde{g}$ from a given plane $z=z'+d$ back to the metasurface plane at $z=z'$. In the scalar case, the reverse propagator is a Fourier propagator taking the form%
\begin{equation}
\label{eq:Phi}
\tilde{\Phi}\left(k_x,k_y; d\right) = \exp\left( -i \sqrt{k^2 - k_x^2 - k_y^2} d\right).
\end{equation}%
Inserting~(\ref{eq:Phi}) into~(\ref{eq:scalar}) yields the complete momentum transformation equation%
\begin{equation}
\tilde{\eta} \left(k_x,k_y\right)= \left[ \tilde{g}\left(k_x,k_y\right) \tilde{\Phi}\left(k_x,k_y; d\right)\right] \ast \tilde{\zeta}\left(k_x,k_y\right). \label{eq:smt}
\end{equation}

The function $\eta(x,y)$ may be considered as a generalization of the optical transfer function (OTF)~\cite{Goodman:96} for a metasurface, albeit without the limitations associated with the OTF, such as normal incidence, large electrical size of the surface and ignoring reflection. Alternatively, the OTF may be considered as the high frequency approximation of $\eta(x,y)$.

\subsection{Vector Transform}
\label{sec:2.2}

A vector field $k$-space basis function set must be chosen to extend the scalar transform to the vectorial transform. This set must satisfy the completeness and orthogonality conditions in order to enable the expansion of any vector field. This expansion allows for the scalarization of the vector problem in a fashion similar to mode matching techniques, where the mode index is the scalar quantity that relates the vector fields involved in the transformation. Once the vector fields on both sides of the metasurface are properly expanded, the momentum transformation follows naturally.

\subsubsection{Momentum Space Basis Functions}

A complete and orthogonal set of vector plane-waves is constructed from Hertz potentials, namely%
\begin{equation}
\label{eq:vpw}
\begin{aligned}
\mathbf{E}^{\text{VPW}} &= \nabla \left[ \nabla \cdot \Pi_e \right] + \frac{\omega^2}{c^2} \Pi_e + i \omega \mu \nabla \times \Pi_h , \\
\mathbf{H}^{\text{VPW}} &= \nabla \left[ \nabla \cdot \Pi_h \right] + \frac{\omega^2}{c^2} \Pi_h - i \omega \epsilon \nabla \times \Pi_e ,
\end{aligned}
\end{equation}%
where $\epsilon$ is the permittivity, $\mu$ the permeability, $c$ the speed of light, $\Pi_{e/h}$ are the electric/magnetic Hertz potentials and the harmonic time dependence $\exp(-i \omega t)$ is assumed with $\omega$ the angular frequency. The choice of the potentials depend on the problem. However, for a $z$-oriented metasurface, and without loss of generality, the potentials are chosen as $\Pi_{e/h} = [A_x^{e/h} \hat{\mathbf{x}} + A_y^{e/h} \hat{\mathbf{y}}] \exp(i \mathbf{k}\cdot\mathbf{r})$, where $\hat{\mathbf{x}}$ and $\hat{\mathbf{y}}$ are the unit vectors in the $x$- and $y$-directions, $\mathbf{r}$ is the position vector, $|\mathbf{k}| = \omega/c$ and $\mathbf{A}^{e/h}$ are arbitrary complex vector amplitudes. Inserting the Hertz potentials in~(\ref{eq:vpw}) yields the electric field components%
\begin{align}
E_x^{\text{VPW}} &= \left[ \left(k^2 - k_x^2\right) A_x^e - k_x k_y A_y^e + k_z A_y^h \omega \mu \right] e^{i \left[ k_x x+ k_y y+k_z z\right]}, \nonumber \\
E_y^{\text{VPW}} &= \left[-k_x k_y A_y^e + \left(k^2 - k_y^2\right) A_y^e -  k_z A_x^h \omega \mu \right] e^{i \left[ k_x x+ k_y y+k_z z\right]}, \nonumber \\
E_z^{\text{VPW}} &= -\left[ k_x k_z A_x^e + k_y k_z A_y^e + \left( k_x A_y^h - k_y A_x^h\right) \omega \mu \right] e^{i \left[ k_x x+ k_y y+k_z z\right]}, \nonumber
\end{align}%
and the magnetic field components%
\begin{align}
H_x^{\text{VPW}} &= \left[ \left(k^2 - k_x^2\right) A_x^h - k_x k_y A_y^h - k_z A_y^e \omega \epsilon \right] e^{i \left[ k_x x+ k_y y+k_z z\right]}, \nonumber \\
H_y^{\text{VPW}} &= \left[-k_x k_y A_y^h + \left(k^2 - k_y^2\right) A_y^h + k_z A_x^e \omega \epsilon \right] e^{i \left[ k_x x+ k_y y+k_z z\right]}, \nonumber \\
H_z^{\text{VPW}} &= -\left[ k_x k_z A_x^h + k_y k_z A_y^h - \left( k_x A_y^e - k_y A_x^e\right) \omega \epsilon \right] e^{i \left[ k_x x+ k_y y+k_z z\right]}. \nonumber
\end{align}

The corresponding completeness and orthogonality follow from the electromagnetic power relation%
\begin{equation}
\label{eq:powerorthogonality}
\iint_{-\infty}^{\infty}{ \hat{\mathbf{z}} \cdot \left[ \mathbf{E}^{\text{VPW}} \left(x,y;\mathbf{k}_z\right) \times \mathbf{H}^{\ast \text{VPW}}\left(x,y;\mathbf{k}_{z'}\right)\right] dx dy} = P_{k_z}^{\text{VPW}} \delta\left(\mathbf{k}_{z'} - \mathbf{k}_{z} \right),
\end{equation}%
where $\mathbf{H}^{\ast}$ denotes the complex conjugate of $\mathbf{H}$ and $P_{k_z}^{\text{VPW}}$ is the power of the vector plane-wave of longitudinal wavevector component $k_z$. Here, it should be noted that the orthogonality of the vector plane-wave set is similar to that of guided modes in the sense that both are power orthogonality relations. Accordingly, the vector plane-wave set forms a momentum space basis set and could be used to expand any vector field in the vicinity of the metasurface. The corresponding transformation pair takes the form%
\begin{align}
&\tilde{\zeta}_{k_z} = \frac{1}{P_{k_z}^{\text{VPW}}} {\iint_{-\infty}^{\infty}{ \mathbf{E}_\perp \left(x,y\right) \times \mathbf{H}_\perp^{\ast \text{VPW}} \left(x,y;k_z\right) d x} d y } \nonumber \\
&\mathbf{E}_\perp \left(x,y\right) = \int_{-\infty}^{\infty}{ \zeta_{k_z} \mathbf{E}_\perp^{\text{VPW}} \left(x,y;k_z\right)  d k_z}, \nonumber
\end{align}
where $k_z$ may be considered as a continuous index for the momentum space modes.

\subsubsection{Reverse Propagator and Transform Equation}

Similar to the scalar case, the reverse propagator is given by%
\begin{equation}
\tilde{\Phi}_{k_z}\left(d\right) = \exp\left( -i k_z d \right), \nonumber \label{eq:Phi_vpw}
\end{equation}
where $k_z$ is directly specified in the vector case and prescribes the transverse wavevector components, in contrast to the scalar case, where $k_z$ is prescribed by the transverse wavevector components. This namely sets the momentum transformation approach for vector fields to be a mode matching approach without violating the wave equation. Using the power orthogonality of the momentum space basis functions~(\ref{eq:powerorthogonality}), the fields in the vicinity of the metasurface are expanded and a momentum transformation relation is established as%
\begin{equation}
\label{eq:vmt}
\tilde{\zeta}_{k_z}^{-} \ast \tilde{\eta}_{k_z} = \tilde{\zeta}_{k_z}^{+} \tilde{\Phi}_{k_z}\left(d\right),
\end{equation}
where the superscripts $-$ and $+$ denote the expanded fields at $z=z'^{-}$ and $z=z'^{+}$, respectively.

Here, it should be noted that the vector nature of the problem introduces an extra level of complication, where the summation over $k_z$ must be taken into account. Nevertheless, such complexity could be viewed as an extra degree of freedom, where the metasurface becomes a vector mode transformer in the $k$-space.

\section{Metasurface Synthesis}
\label{sec:3}

Metasurfaces are formally described as two-dimensional discontinuities in the electromagnetic field~\cite{Idemen:11}. Their reduced dimensionality grants them a great potential for controlling electromagnetic fields without the disadvantages of excessive loss, weight and fabrication difficulty associated with three-dimensional metamaterials~\cite{Kuester:03,Holloway:05,Pfeiffer:13a}. In~\cite{Holloway:12}, the modeling process is carried out by determining the susceptibility dyadics for given fields on both sides of the metasurface. Here, the action of a metasurface is approached under the perspective of the transformation of the momentum of the incident field into that of the transmitted field. Accordingly, metasurface synthesis is also approached under the perspective of light wave momenta, as opposed to susceptibilities. This means that the design of metasurface scattering elements directly relates their geometrical degrees of freedom to their effect on the wave momentum.

Modeling of metasurface elements might be analytical or numerical. In either case, a set of scattering elements must be chosen such that their momentum response covers the range of momentum transformation of interest. This ensures that indeed the metasurface is capable of the required momentum transformation between the incident, reflected and transmitted fields. For this purpose, a set of basis scattering elements are chosen, such that their domain of degrees of freedom maps to the range of momentum change of interest. The mapping between the domain of degrees of freedom of an isolated scattering element to its action on the wavevector components is established by way of analytic modeling or full-wave simulation as done in~\cite{Yu:11}. The resulting maps are then used as lookup maps to identify the suitable element parameters for a specific transformation operation. The elements are then combined to form the complete metasurface. The construction of such lookup maps is beyond the scope of this paper, whose main objective is to establish the momentum transformation method, rather than providing implementation details.

\section{Illustrative Examples}
\label{sec:4}

In this section, two types of momentum transformation related to field radiation at distance are demonstrated. The first type is manipulation in the Fresnel (near-field) region, and the second is manipulation in the Fraunhofer (far-field) region. Conventionally, each radiation region is associated with different dominant field behaviors, for instance, in the Fresnel region, the physical geometry of the metasurface is identifiable and has thus a direct impact on the field distribution, whereas in the Fraunhofer region, the metasurface is assumed to be far enough such that its geometry is indistinguishable.

\subsection{Manipulation in Fresnel Region}
\label{sec:4.1}
In the Fresnel region, diffraction patterns typically significantly differ from that observed at large distance (infinity) away from the source, and vary significantly with the observer position with respect to the source. In this region, all (polarized) field components have to be taken into account and non-radiative field components are expected to exist.

Two examples of field manipulation in the Fresnel region are presented. The first example is the conversion of an incident Gaussian beam into a vortex hypergeometric-Gaussian beam, and the second is the generation of a ``delayed-start'' accelerating Airy beam from an incident plane-wave.

\subsubsection{Vortex Hypergeometric-Gaussian Beam}

Hypergeometric-Gaussian (HyG) beams are a class of paraxial waves solutions with a complex amplitude that is proportional to a confluent hypergeometric function. The beam intensity profile is characterized by a single brilliant ring with the field amplitude vanishing at its center. HyG beams carry topological charges, i.e. their phasefronts are helical, and thus have a singular phase profile. In quantum mechanics, HyG beams describe the eigenfunctions of the photon orbital angular momentum, and since they carry orbital angular momentum, they are of interest in applications, such as micro- and nano-particle manipulation and orbital angular momentum multiplexing~\cite{Kotlyar:08}.

The HyG beam field is defined as%
\begin{equation}
\label{eq:HyG}
\psi_{\text{HyG}}\left(\rho,\phi,z\right) = \frac{\Gamma\left(1+|m|+\frac{p}{2}\right)}{\Gamma\left( |m|+1\right)} \frac{i^{|m|+1}\chi^{|m|/2}\xi^{p/2}}{\left[\xi + i\right]^{1+|m|/2+p/2}}  e^{i m \phi - i \chi} {}_1F_1 \left(-\frac{p}{2},|m|+1;\frac{\chi \left[ \xi+i\right]}{\xi \left[\xi-i\right]} \right),
\end{equation}%
where $\Gamma(x)$ is the gamma function, ${}_1F_1(a,b;x)$ is the confluent hypergeometric function, $w_0$ is the beam waist, $p \geq - |m|$ is a real valued parameter, $m$ is the topological charge, $p$ and $m$ together specify the HyG mode, $\chi = \rho^2/(w_0^2 [\xi+i])$, and $\xi = z/z_{\text{R}}$ with $z_{\text{R}} = \pi w_0^2/\lambda$ the Rayleigh range.

HyG beams are unstable below the Rayleigh range, $z_{\text{R}}$, meaning that one mode can degenerate into others. Hence it is of interest to generate the HyG beam directly beyond the Rayleigh range in order to avoid the inherent beam instability. In this example, the HyG beam is generated by illuminating the metasurface by an ordinary Gaussian beam,%
\begin{equation}
\label{eq:GB}
\psi_{\text{GB}} \left(\rho, \phi, z \right) = \frac{w_0}{w\left(z\right)} \exp\left(-\frac{r^2}{w\left(z\right)^2} - i k \left[ z + \frac{r^2}{2 R\left(z\right)} \right] - i \vartheta\left(z\right)  \right),
\end{equation}%
where $r^2 = \rho^2+z^2$, $R(z)=z[1+(z_{\text{R}}/z)^2]$ is the beam radius of curvature, $w(z)=w_0 [1+(z/z_{\text{R}})^2]^{1/2}$ is the beam waist with $w_0$ the beam waist at $z=0$, and $\vartheta(z) = \arctan(z/z_{\text{R}})$ is the Gouy phase. The $k$-space representation of the field expressions in~(\ref{eq:HyG}) and~(\ref{eq:GB}) are substituted into~(\ref{eq:smt}), where the reverse propagator in this particular example is used as an advance propagator, i.e. it translates the beam profile in the negative $z$-direction as opposed to the conventional positive translation. The purpose of the propagator here is to generate the HyG beam at the plane of the metasurface with the characteristics of a HyG at $z = z_{\text{R}}$, thus avoiding the instability region of the beam. The propagator in this case is then given by%
\begin{equation}
\tilde{\Phi} = \exp\left(i \sqrt{k^2 - k_x^2 - k_y^2} z_{\text{R}} \right). \nonumber
\end{equation}

\begin{figure*}[tpb]
\centering
\subfloat[]{
	\label{fig:2a}
	\includegraphics[width=2.5in]{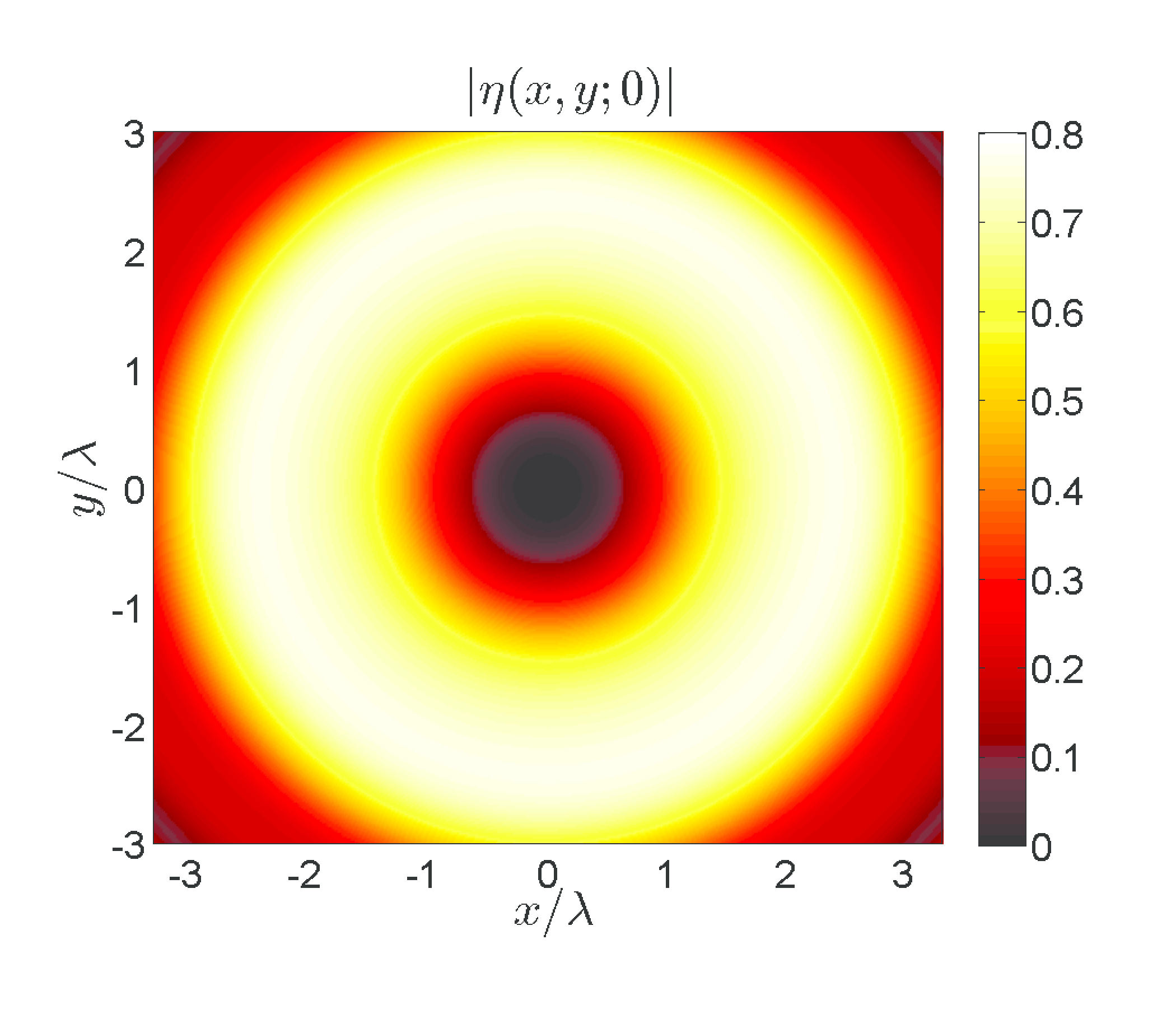}
}
\subfloat[]{
	\label{fig:2b}
	\includegraphics[width=2.5in]{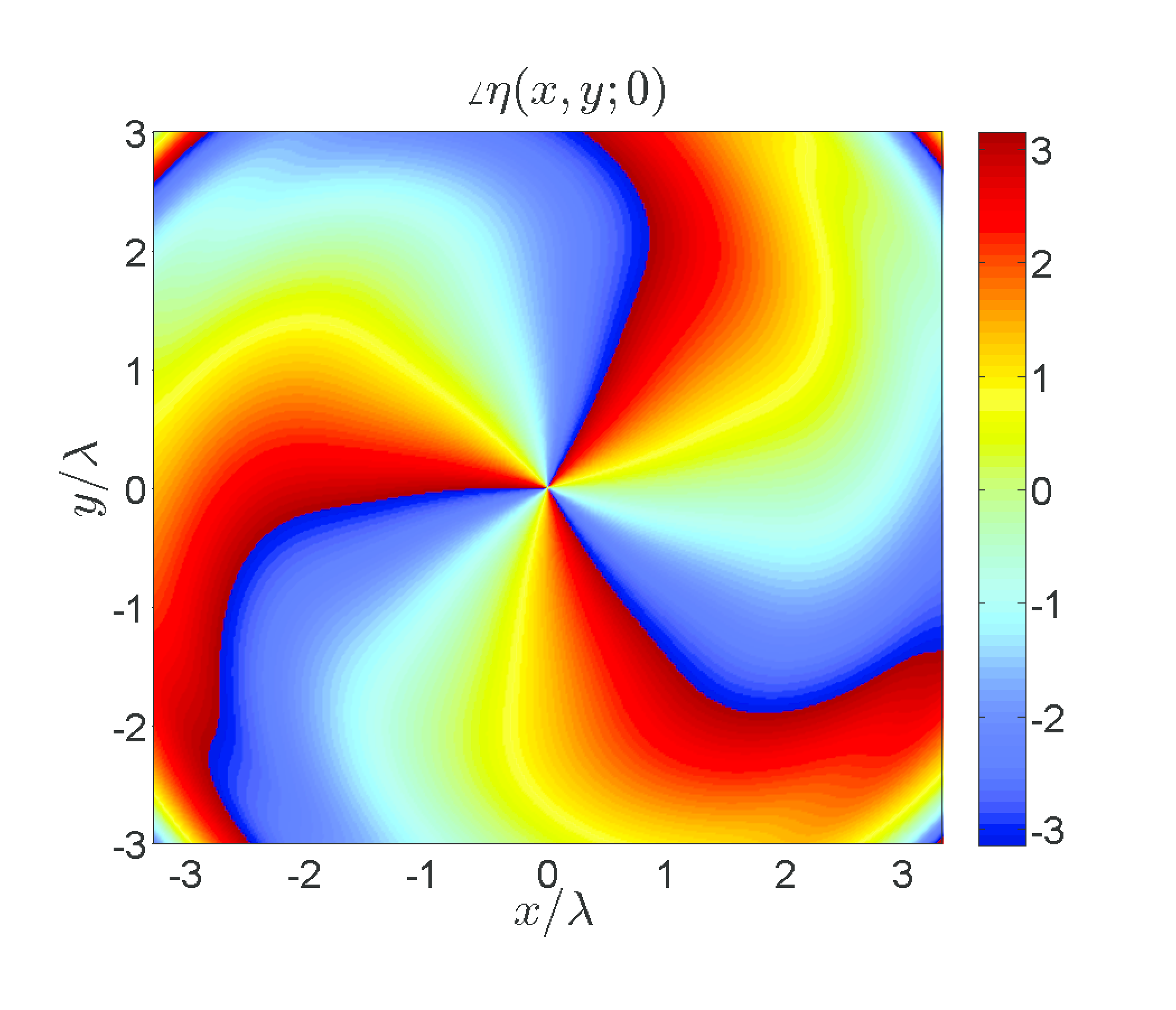}
}
\\
\subfloat[]{
	\label{fig:2c}
	\includegraphics[width=2.5in]{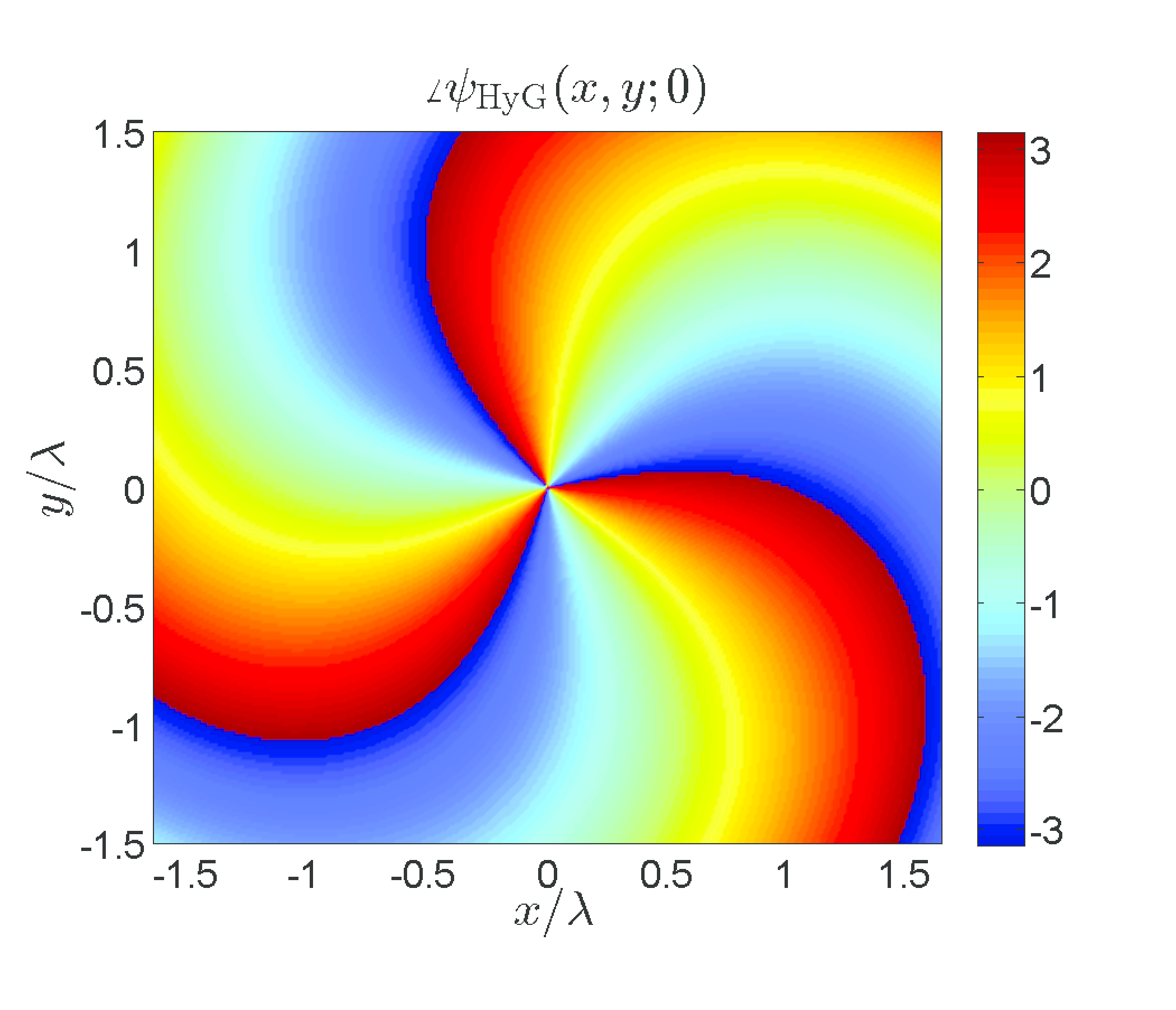}
}
\subfloat[]{
	\label{fig:2d}
	\includegraphics[width=2.5in]{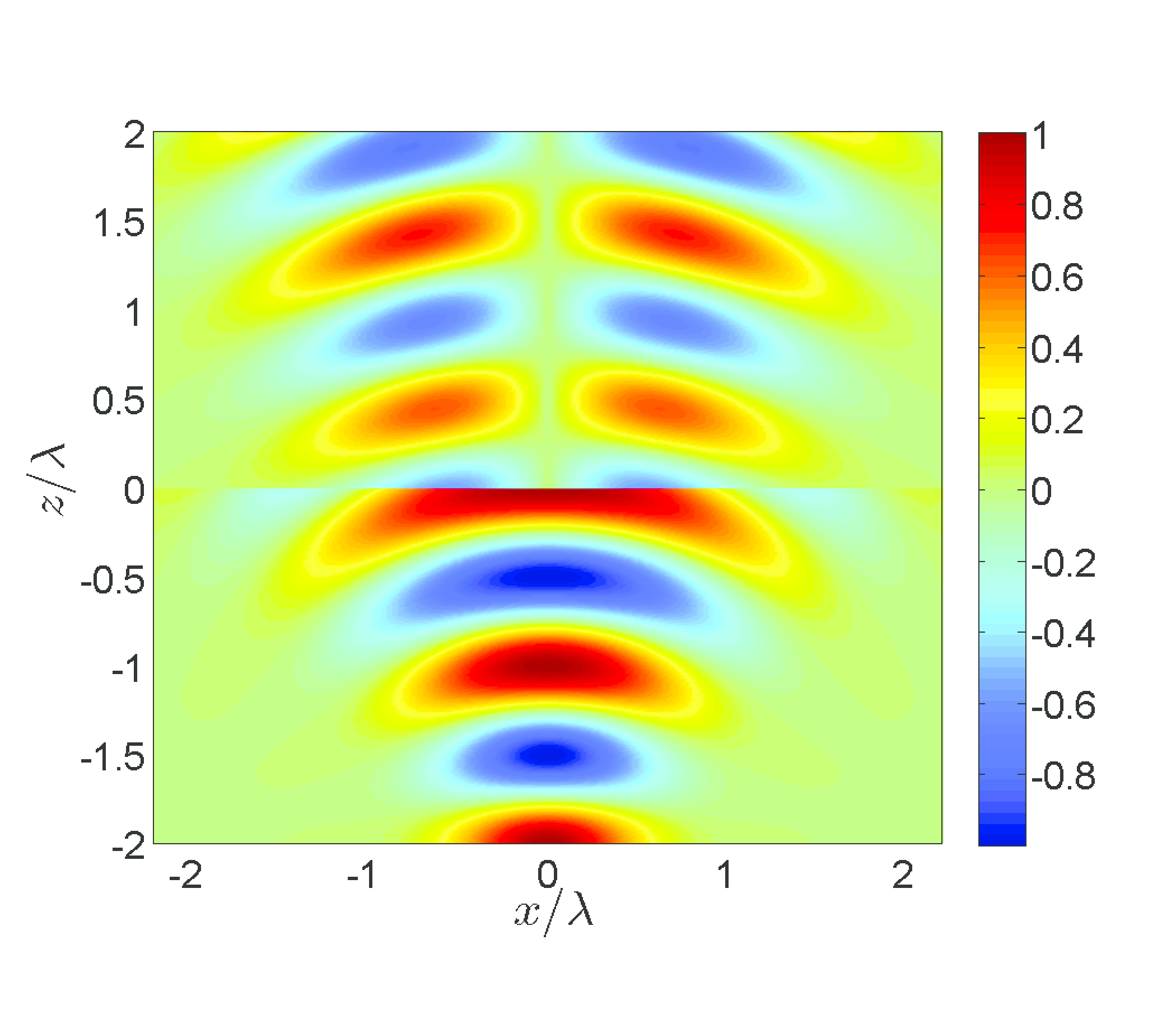}
}
\caption{Generation of a hypergeometric-Guassian beam from an ordinary Gaussian beam. The amplitude and phase of the direct space metasurface function at $z=0$ are plotted in \protect\subref{fig:2a} and \protect\subref{fig:2b}. The phase of the transmitted field is plotted in \protect\subref{fig:3c} showing a vorticity corresponding to a topological charge $m=3$. A propagation plot in the $y=0$ plane is presented in \protect\subref{fig:3d} showing the incident Gaussian beam and the transmitted vortex hypergeometric-Gaussian beam.}
\label{fig:2}
\end{figure*}

Figure~\ref{fig:2} shows the metasurface function required to convert an ordinary Gaussian beam into a vortex hypergeometric-Gaussian beam. Figures~\ref{fig:2}\subref{fig:2a} and \ref{fig:2}\subref{fig:2b} show the magnitude and phase functions, respectively, of the metasurface that convert a Gaussian beam with beam waist $w_0 = \lambda$ and zero Guoy phase into a vortex hypergeometric-Gaussian beam with the same beam waist, a topological charge $m=3$, and $p=1$. The transmitted hypergeometric-Gaussian beam is advance-propagated to distance $z=z_{\text{R}}$ to avoid the beam instability range. The phase of the vortex beam at the metasurface plane is plotted in Fig.~\ref{fig:2}\subref{fig:2c}, and a plot of the incident Gaussian beam and the transmitted vortex hypergeometric-Gaussian beam after the metasurface in the $y=0$ plane is presented in Fig.~\ref{fig:2}\subref{fig:2d}.

\subsubsection{Delayed-Start Accelerating Beam}

Solutions to the paraxial wave equation (Schr\"{o}dinger equation) in terms of Airy functions were analyzed in the context of quantum mechanics in \cite{Berry:79}. Airy beams possess asymmetric parity about the origin and follow parabolic propagation trajectories. The curvature associated with the lateral shift was the initial reason for describing Airy beams as accelerating beams. An Airy beam is defined by \cite{Siviloglou:07}%
\begin{align}
\label{eq:AiB}
\psi_{\text{AiB}} \left(x,z \right) &= \text{Ai}\left(s - \frac{\zeta^2}{4} \right) \exp\left( i \left[\frac{s\zeta}{2} - \frac{\zeta^3}{12} \right] \right),
\end{align}%
where $\text{Ai}(x)$ is the Airy function, $s = x/x_0$, $\zeta = z/kx_0^2$, and $x_0$ is a arbitrary scaling factor.

As may be seen in~(\ref{eq:AiB}), Airy beams start their parabolic trajectory (acceleration) at $z=0$. In order to delay the start of the trajectory curving to a distance $z=d$, the reverse propagator is applied to translate the zero acceleration plane from $z=0$ to $z=d$. In this example, a normal incident plane-wave with $\psi_{\text{PW}} = \exp(i k z)$ illuminates the metasurface and an Airy beam starting its parabolic trajectory at $z=d$ is generated. The relevant reverse propagator is%
\begin{equation}
\label{eq:prop}
\tilde{\Phi} = \exp\left(-i \sqrt{k^2 - k_x^2} d \right). \nonumber
\end{equation}
Here it should be noted that neither the incident nor the transmitted fields have $y$-directed momentum components.

\begin{figure*}[tpb]
\centering
\subfloat[]{
	\label{fig:3a}
	\includegraphics[width=2.5in]{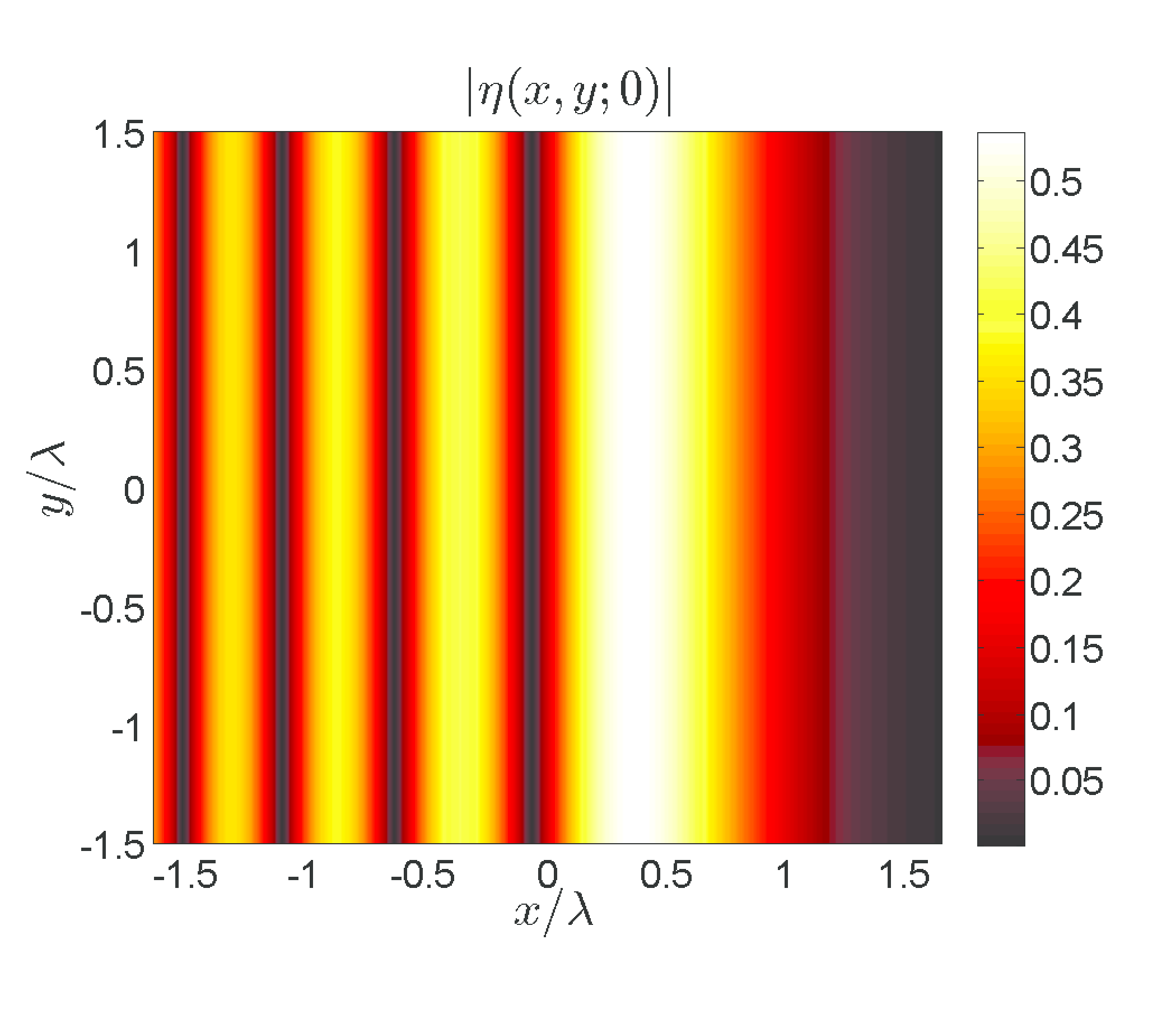}
}
\subfloat[]{
	\label{fig:3b}
	\includegraphics[width=2.5in]{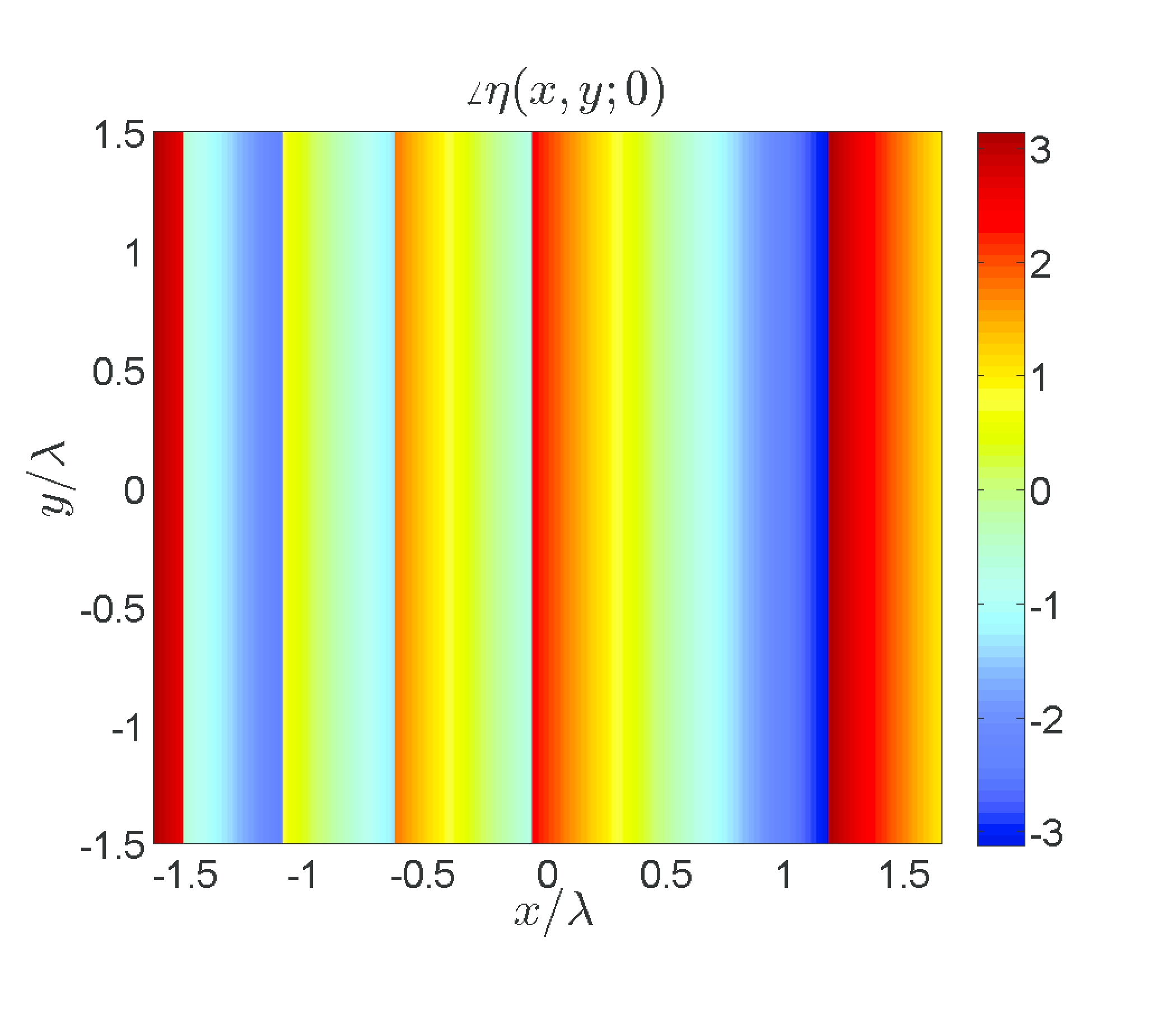}
}
\\
\subfloat[]{
	\label{fig:3c}
	\includegraphics[width=2.5in]{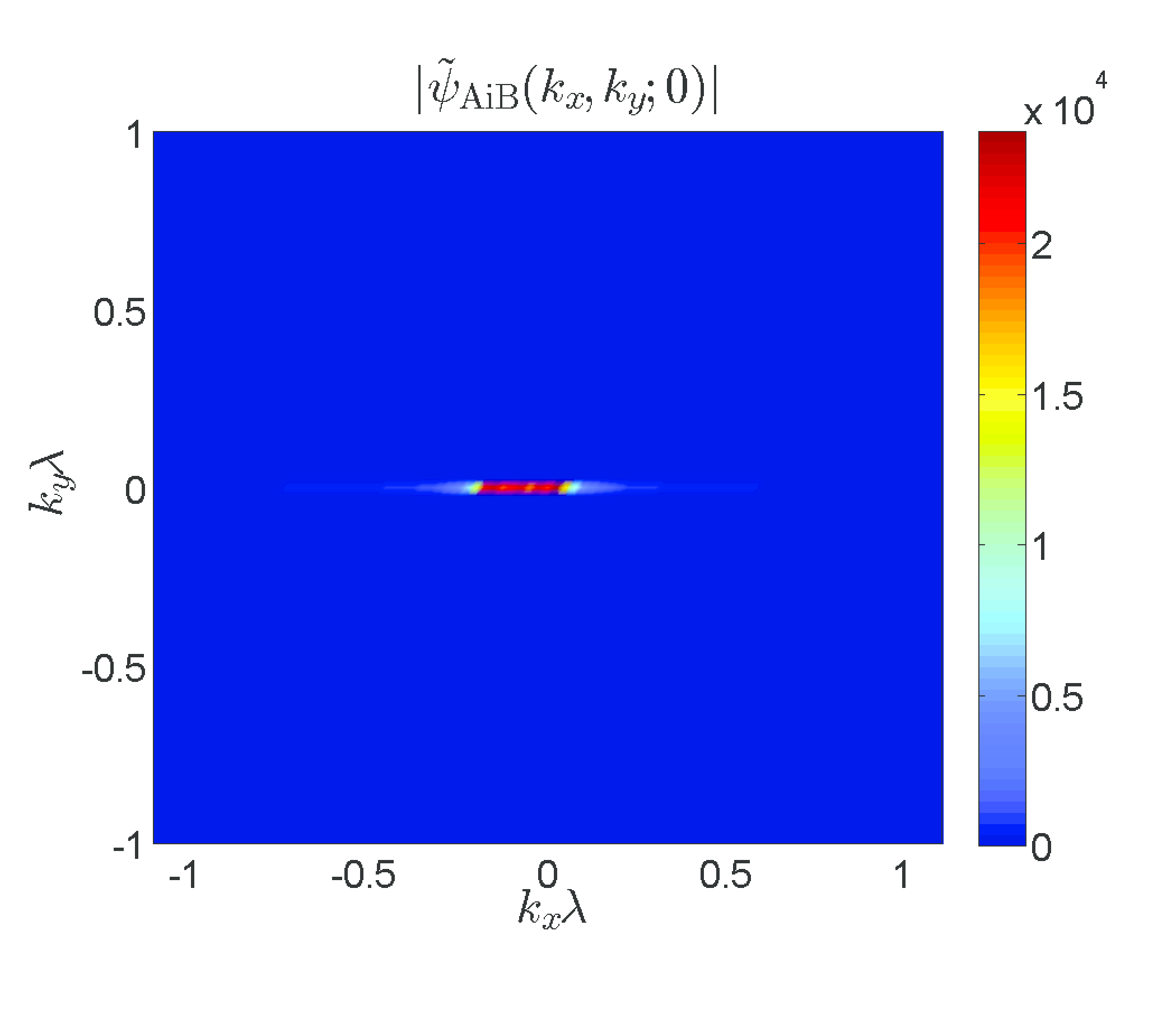}
}
\subfloat[]{
	\label{fig:3d}
	\includegraphics[width=2.5in]{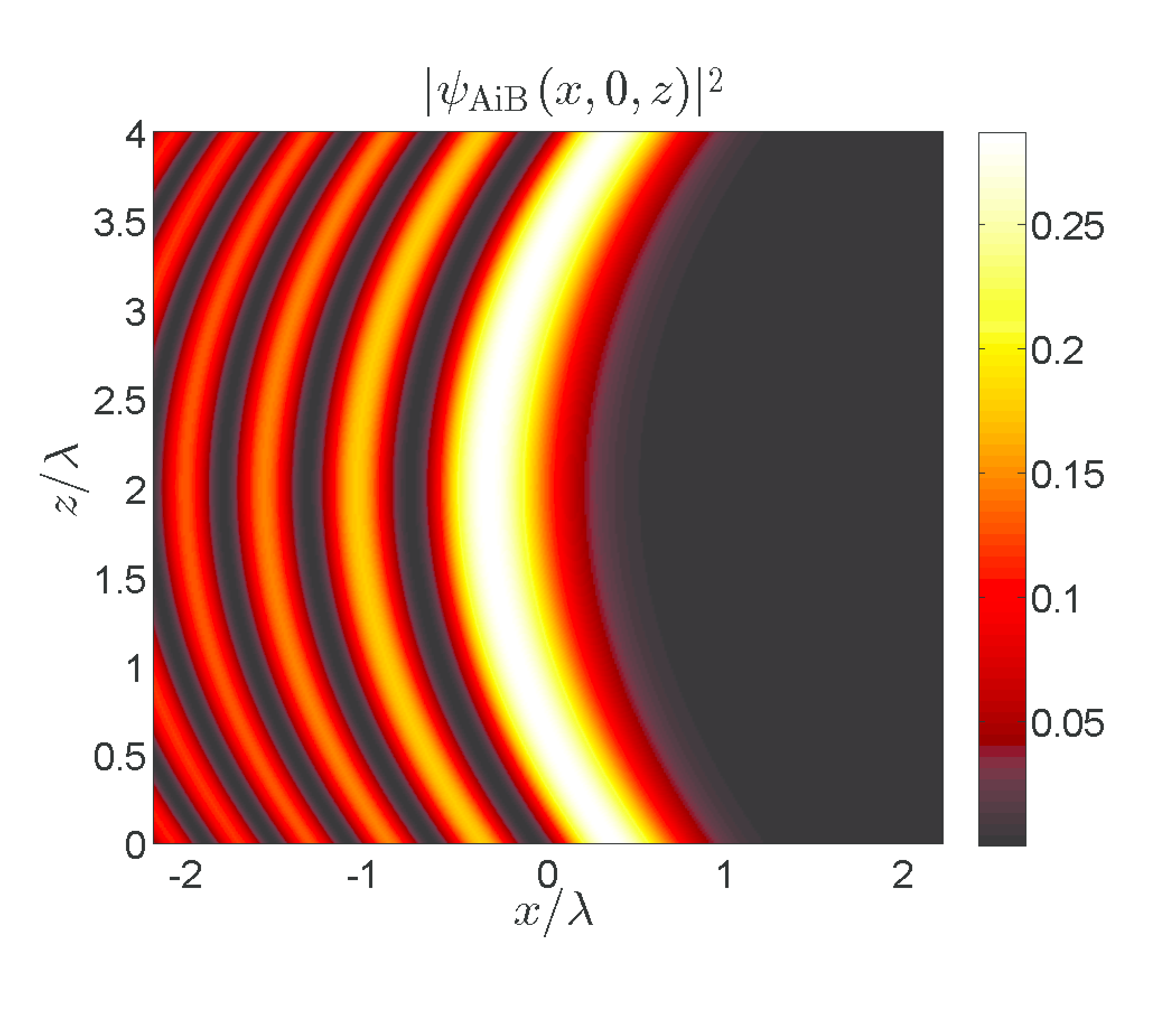}
}
\caption{Generation of a ``delayed-start'' accelerating beam from a plane-wave. The amplitude and phase of the direct space metasurface function at $z=0$ are plotted in \protect\subref{fig:3a} and \protect\subref{fig:3b}, respectively. The $k$-space representation of the generated delayed-start Airy beam is plotted in \protect\subref{fig:3c} and the intensity profile of the beam in the $x-z$ plane is plotted in \protect\subref{fig:3d}. The beam starts its curvature towards the positive $x$-axis at $z = 2\lambda$.}
\label{fig:3}
\end{figure*}

Figure~\ref{fig:3} shows the metasurface function generating a delayed-start Airy beam. In Figs.~\ref{fig:3}\subref{fig:3a} and \ref{fig:3}\subref{fig:3b}, the amplitude and phase values of the metasurface function in direct space are plotted. Illuminating this metasurface with a normally incident plane-wave with wavelength $\lambda$ generates a delayed-start Airy beam with $x_0 = \lambda/3$ and positive curvature start at $z=2\lambda$. In Fig.~\ref{fig:3}\subref{fig:3c} the momentum space representation of the Airy beam is plotted, and its intensity profile in the $x-z$ plane is plotted in Fig.~\ref{fig:3}\subref{fig:3d}.

\subsection{Manipulation in Fraunhofer Region}
\label{sec:4.2}

The main characteristic of electromagnetic fields in the Fraunhofer region is that the electric field and the magnetic field form a transverse electromagnetic wave. Moreover, the fields in this region are polarized, therefore only one component of each field exist in this region.

Two examples of field manipulation in the Fraunhofer region are presented. The first example is the generation of a pencil beam, and the second is the demonstration of a holographic repeater, where the source field is reconstructed at a remote distance.

\subsubsection{Pencil Beam Radiator}

A pencil beam is a beam of electromagnetic radiation in the form of a narrow cone. Antennas which are strongly directive in both azimuth and elevation, such as phased array antennas, are often used to generate pencil beams. However, such phased array antennas suffer from several disadvantages including the need for complex and expensive components and circuitry. Alternatively, resorting to the super gain concept~\cite{Balanis:05}, the same pencil beam may be, in principle, achieved by an antenna with a single feed. In conventional antennas, however, super gain demands rapid current variations along the antenna structure, which severely limits the realizability due to huge ohmic loss and excessive tolerance sensitivity~\cite{Balanis:05}.

Metasurfaces offer a decisive advantage in the generation of pencil beams, since they do not require variable actual current sources. Accordingly, super gain is achievable using momentum transformation without the issues encountered in conventional super gain antennas. Metasurface implementations offer great flexibility since the amount of induced momentum variation by any individual scattering element may be determined independent from its neighboring elements. The physical limitation in the case of metasurface implementation is the requirement that the size of the metasurface scattering element has to be considerably smaller than the operational wavelength. Following Huygen's principle, the metasurface in this case acts as an equivalent antenna array. The corresponding equivalent array factor for the required pencil beam formation is derived in straightforward fashion using standard array synthesis techniques \cite{Balanis:05}. The required beam solid angle, which is the figure of merit for a pencil beam is directly computed from the equivalent array factor, and reads in the spherical coordinates $(r,\theta,\phi)$,%
\begin{equation}
\label{eq:solidangle}
\Delta \Omega = 2 \pi \int_0^\pi {\frac{|A\left(\theta\right)|^2}{|A\left(\theta_t\right)|^2} \sin\left(\theta\right) d \theta},
\end{equation}
where $A(\theta)$ is the array factor, $\theta_t$ is the angle at which the beam maximum occurs. Following~(\ref{eq:solidangle}), the array factor is transformed to the $k$-space and used for momentum transformation.

\begin{figure*}[tpb]
\centering
\subfloat[]{
	\label{fig:4a}
	\includegraphics[width=2.5in]{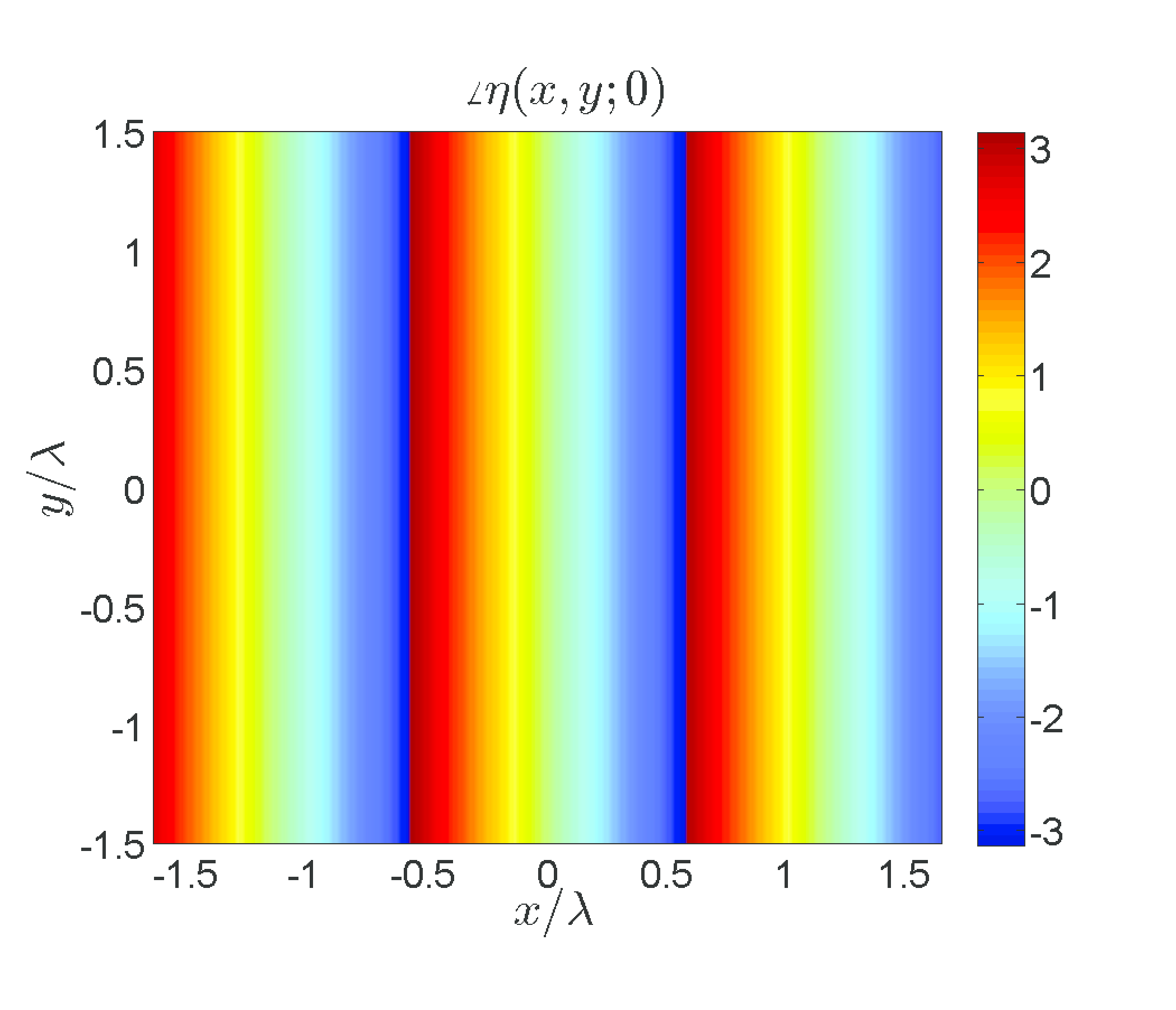}
}
\subfloat[]{
	\label{fig:4b}
	\includegraphics[width=2.5in]{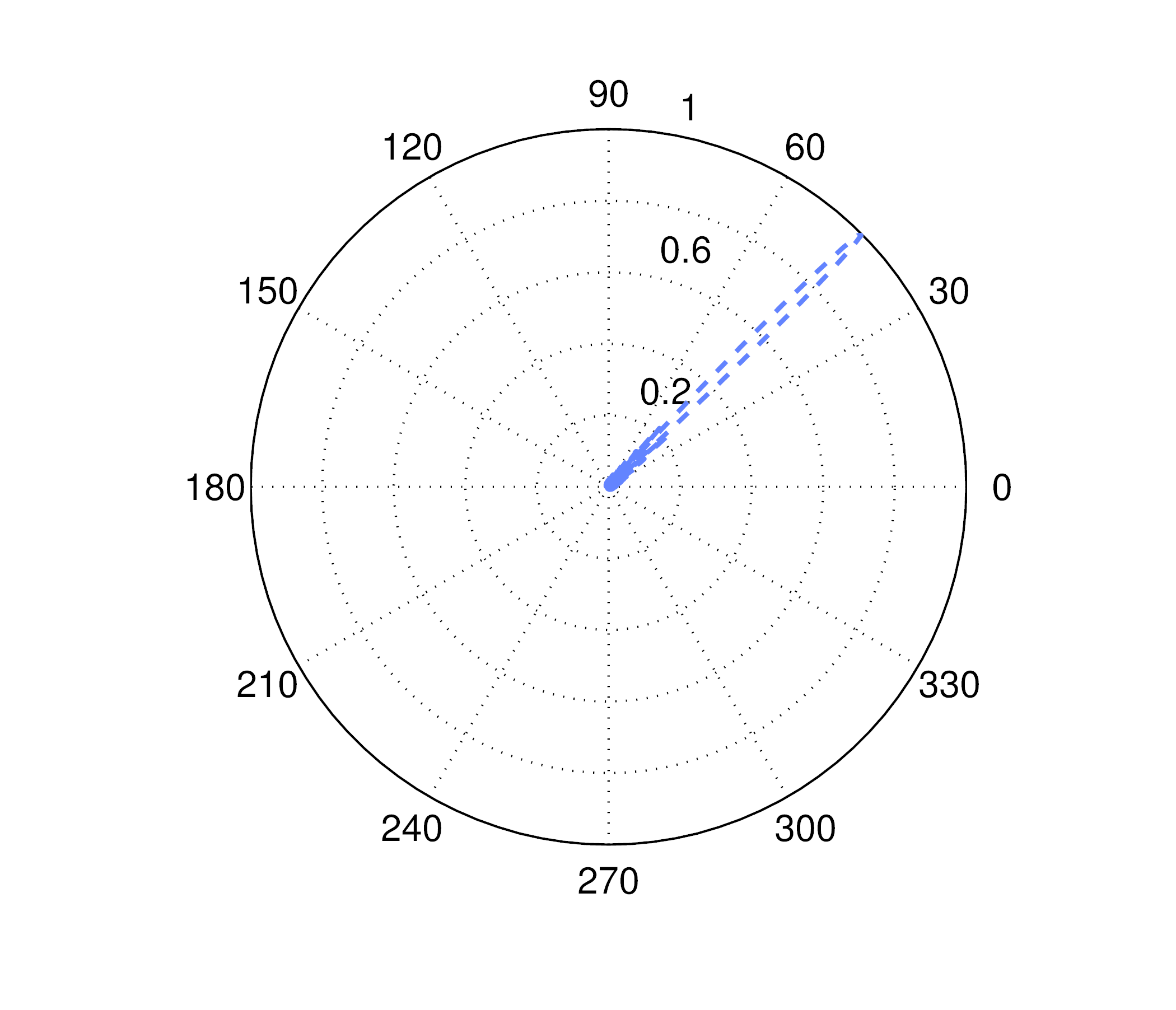}
}
\caption{Generation of a tilted pencil beam from a plane wave. The phase function of the metasurface to generate the tilted pencil beam at angle $\theta_t = \pi/4$ is depicted in \protect\subref{fig:4a}. The equivalent directivity of the metasurface is plotted in \protect\subref{fig:4b}.}
\label{fig:4}
\end{figure*}

Figure~\ref{fig:4} presents an example of a tilted pencil radiation from a metasurface illuminated by a normally incident plane-wave with wavelength $\lambda$ and polarized along $y$-direction. The radiated pencil beam is tilted by an angle $\theta_t = \pi/4$ measured from the metasurface axis along the $x$-direction and polarized along $y$-direction. The metasurface acts as a transmitting phased array antenna (transmit-array) and only modifies the local phases of the incident plane-wave. The metasurface function is determined using~\eqref{eq:vmt} under the assumption of no reflection. The phase of the metasurface function is plotted in Fig.~\ref{fig:4}\subref{fig:4a} and the equivalent antenna array pattern is plotted in Fig.~\ref{fig:4}\subref{fig:4b}. The equivalent maximum directivity of this metasurface is $35.12 \,\mathrm{dB}$ and the beam width is $\Delta \Omega \approx \pi/64$.

\subsubsection{Holographic Repeater}

A holographic repeater is a device that reconstructs an image of a primary source at a desired distance away from that source. There is significant interest in such repeaters, particularly in point-to-point free-space communication applications. Previous research work in this area employed near-field patterned plates~\cite{Grbic:08} or antenna arrays and a metascreen~\cite{Markley:11} to replicate the source image. Here, a holographic repeater employing a single metasurface is demonstrated.

The main difficulty in reconstructing a high fidelity source image is that far away from the source, the evanescent components of the source field rapidly decay thus preventing perfect reconstruction. A perfect source reconstruction based on superlenses made out of doubly-negative metamaterials was proposed in~\cite{Pendry:00}. Here, an equivalent two-dimensional perfect lens is constructed by altering the momentum direction of the evanescent components such that the source information is preserved and may be perfectly reconstructed at a distance.

Here, a two-dimensional holographic image of a radiating point source is reconstructed at a distance $z=d$ away from the metasurface. It should be understood that faithful field recovery is possible only for $z\geq d$, since causal reconstruction of the fields radiating in the negative $z$-direction would require sources at $z>d$. Assuming a plane-wave illumination of the form $\exp(i k z)$, the metasurface function of the holographic repeater is found as%
\begin{equation}
\label{eq:repeater}
\tilde{\eta}\left(k_x,k_y\right) = \beta \exp(-i z_d \sqrt{k^2 - \left[ k_x^2 + k_y^2 \right]}), \nonumber
\end{equation}
where $\beta$ is an arbitrary constant.

\begin{figure*}[tpb]
\centering
\subfloat[]{
	\label{fig:5a}
	\includegraphics[width=2.5in]{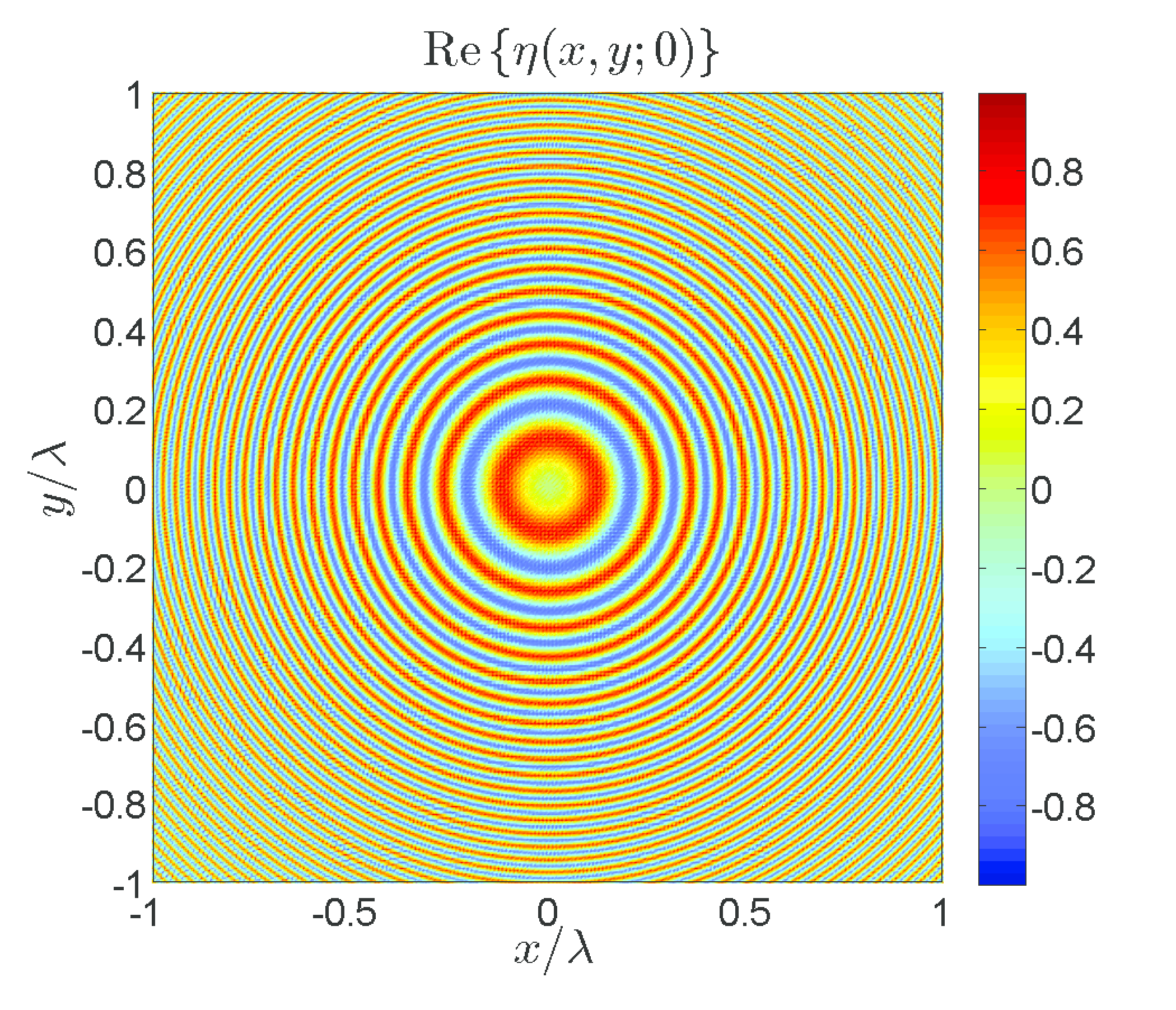}
}
\subfloat[]{
	\label{fig:5b}
	\includegraphics[width=2.5in]{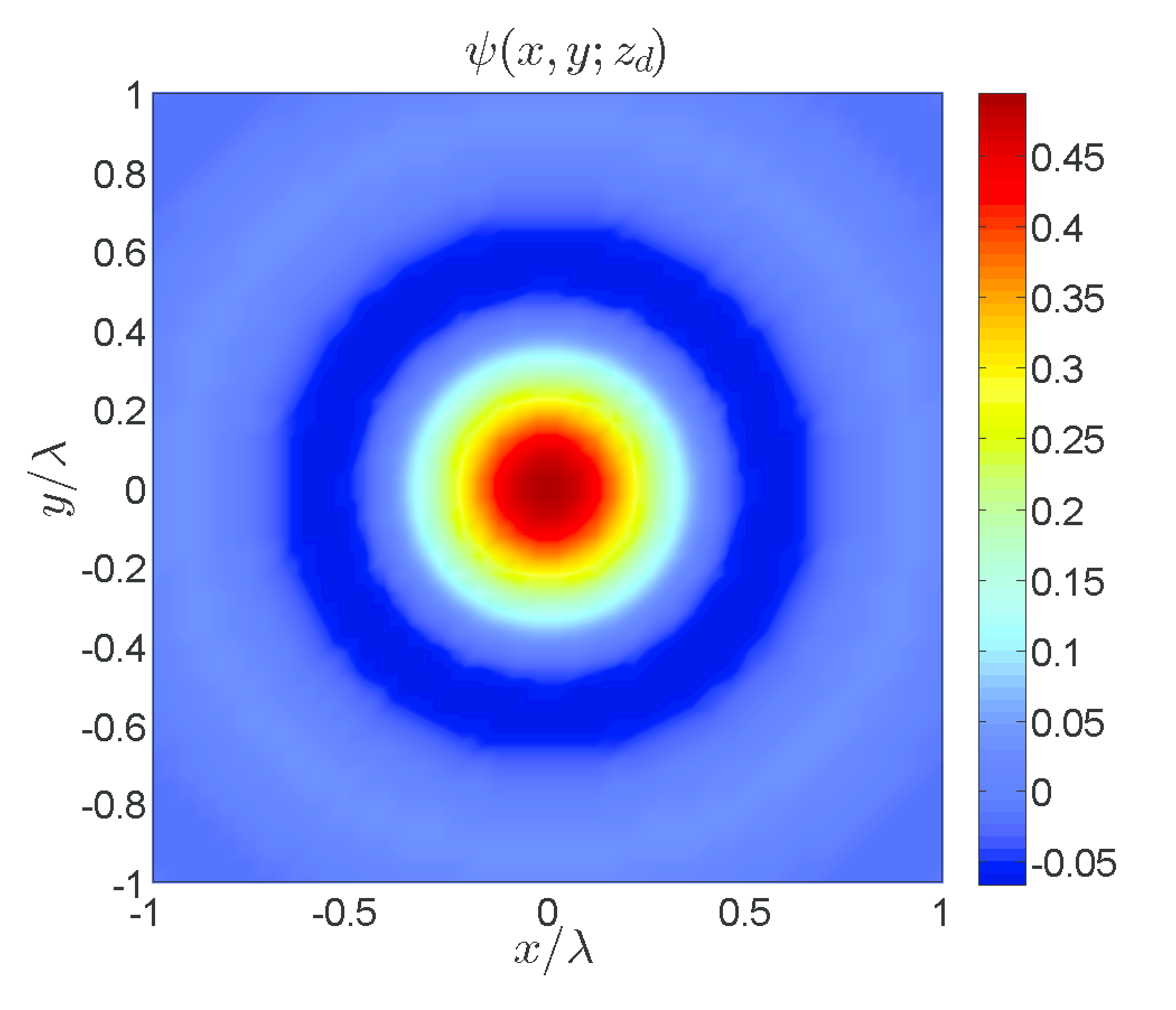}
}
\caption{Holographic repetition from a plane-wave. The direct space metasurface function at $z=0$ is plotted in \protect\subref{fig:5a} and the reconstructed field at $z=300\lambda$ due to a normally incident plane-wave is plotted in \protect\subref{fig:5b}. The metasurface function bears an uncanny resemblance to a Fresnel lens.}
\label{fig:5}
\end{figure*}

Figures~\ref{fig:5}\subref{fig:5a} and \ref{fig:5}\subref{fig:5b} respectively depict the metasurface function and the resulting reconstructed field of unit electrical strength source due to an incident plane-wave with wavelength $\lambda$ and a reconstruction distance $z_d=300 \lambda$. The metasurface is assumed to have dimensions of $20 \lambda \times 20 \lambda$ and its function resembles that of Fresnel lenses, as expected. The reconstructed field does not perfectly recover the source field due to the finite size of the metasurface.

\section{Conclusions}
\label{sec:5}

A generalized framework for light manipulation at distance using momentum transformation is introduced. Two types of transformation relations were derived for scalar and vector field cases, respectively. In addition, the concept of the reverse propagator was incorporated in the synthesis process to facilitate manipulation of the light wave at a distance away from the metasurface. The momentum transformation equations yield a metasurface description in the $k$-space, which may be transformed back into direct space in a straightforward fashion. The direct space description may then be physically implemented using element lookup map techniques. Examples of light manipulation demonstrating Fresnel zone (near-field) and Fraunhofer zone (far-field) effects were shown.

The proposed momentum transformation method offers a direct, flexible and systematic approach to manipulate waves at distance. It avoids the convolved computation of electric and magnetic susceptibility dyadics, since all computations are performed as straightforward momentum variations in the $k$-space. The mapping between the metasurface description in the $k$-space and the physical realization is a straightforward process via scattering element lookup maps, which may be constructed through electromagnetic analysis or full-wave simulation. The momentum transformation approach is also flexible due to its generality, in the sense that it can accommodate scalar and vector field transformations. An additional level of flexibility follows from the synthesis process, which does not require prior knowledge of field distributions at any intermediate distance between the metasurface and the destination plane. Finally, synthesizing metasurfaces based on momentum transformation is a rigorous, systematic and unified approach for treatment of all cases of wave manipulation at distance. This avoids the serious of trial-and-error based approaches, and also avoids resorting to case-specific synthesis techniques.

Future work may include investigating the synthesis of multi-layered metasurfaces, where several metasurfaces are stacked back-to-back, and active metasurfaces and non-Foster type metasurfaces. Such structures are expected to provide greater flexibility in terms of the momentum variation they can induce, in addition to providing larger bandwidth.

\section*{Appendix: Linear-Shift Variance of the Metasurface-Field Interaction in Direct Space}
A formal representation of~(\ref{eq:scalar}) is derived applying the distribution theory for the momentum discontinuity in the metasurface plane. First, the difference between the fields at the two sides of the metasurface is considered%
\begin{align}
g\left(x,y \right) - f\left(x,y\right)  &= \Lambda\left(x,y\right), \nonumber \\
\tilde{g}\left(k_x,k_y \right) - \tilde{f}\left(k_x,k_y\right)  &= \tilde{\Lambda}\left(k_x,k_y\right), \label{eq:lambda} 
\end{align}
where $\Lambda(x,y)$ is the field difference. Equation~(\ref{eq:lambda}) seems a priori contradictory, since $\tilde{\Lambda}(k_x,k_y)$ has to be zero due to the momentum conservation. However, this apparent contradiction vanishes when the momenta are considered in the sense of distributions, i.e. generalized functions describing the physics. This may be shown by using the Fourier transform property of distributions \cite{Strichartz:03}%
\begin{equation}
\label{eq:distrib}
\langle \mathcal{F}^{-1}\left\{ \tilde{\Lambda} \right\}, \tilde{\varphi} \rangle = \langle \tilde{\Lambda}, \mathcal{F}^{-1}\left\{ \tilde{\varphi} \right\} \rangle,
\end{equation}
where $\mathcal{F}^{-1}\{\tilde{\psi}\}$ is the inverse Fourier transform of $\tilde{\psi}$, $\tilde{\varphi}$ is a well-behaved, i.e. smooth and has a compact support, testing function. This well-behavior of $\tilde{\varphi}$ is also necessary when describing causal physical phenomena. Hence,%
\begin{equation}
\langle \psi(x,y), \varphi(x,y) \rangle = \iint_{-\infty}^{\infty} { \psi(x,y) \varphi(x,y) dx dy }. \nonumber
\end{equation}
Choosing the testing function to be the general transfer function of the metasurface, and considering the momentum conservation relation, (\ref{eq:distrib}) reduces to%
\begin{equation}
\langle \mathcal{F}^{-1}\left\{\tilde{\Lambda}\right\} , \tilde{h} \rangle = \langle 0, \mathcal{F}^{-1}\left\{\tilde{h}\right\} \rangle, \nonumber
\end{equation}
which requires $\text{supp}[\tilde{\Lambda}] = 0$, i.e. $\tilde{\Lambda} = \tilde{\alpha}(k_x,k_y)\delta(k_x,k_y)$, with $\tilde{\alpha}$ an arbitrary spectral function. Inserting this relation into~(\ref{eq:meta}) and changing the order of the integrals, assuming the well-behavior of $h$, recovers~(\ref{eq:lsv}) by choosing $\alpha(x,y) = [\eta(x,y) - 1] f(x,y)$.

\end{document}